\documentclass[conference]{IEEEtran}

\ifCLASSINFOpdf

\else
 
\fi

\usepackage{graphicx}
\DeclareGraphicsExtensions{.eps}
\usepackage{caption}
\usepackage{subcaption}
\usepackage{cite}

\usepackage{graphicx}
\DeclareGraphicsExtensions{.eps}
\usepackage{setspace}
\providecommand{\PSforPDF}[1]{#1}
\usepackage[cmex10]{amsmath}
\usepackage{amsfonts,amssymb,amsthm}
\usepackage{times}
\PSforPDF{\usepackage{psfrag}}
\usepackage{algorithmic}
\usepackage{algorithm}
\usepackage{array}
\usepackage{multirow}
\usepackage{comment}
\usepackage{balance}
\usepackage{mdwmath}

\usepackage{mdwtab}
\usepackage{eqparbox}
\usepackage{nicefrac}
\usepackage{url}

\usepackage{exscale,relsize}
\usepackage{color} 
\begin{document}

\title{Dimensioning of PA for massive MIMO system with load adaptive number of antennas}

\author{\IEEEauthorblockN{M M Aftab Hossain}
\IEEEauthorblockA{School of Electrical Engineering\\
Aalto University, Finland\\
Email: mm.hossain@aalto.fi}
\and
\IEEEauthorblockN{Riku J\"antti}
\IEEEauthorblockA{School of Electrical Engineering\\
Aalto University, Finland\\
Email:riku.jantti@aalto.fi}
\and
\IEEEauthorblockN{Cicek Cavdar}
Wireless@KTH, \\
KTH Royal Institute of Technology, Sweden\\
Email:cavdar@kth.se}
\maketitle

\begin{abstract}
This paper takes into consideration the non-ideal efficiency characteristics of realistic power amplifiers  (PAs) along with the daily traffic profile in order to investigate the impact of PA dimensioning on the energy efficiency (EE) of load adaptive massive MIMO system. A multicellular system has been considered where each base station (BS) is equipped with a large number of antennas to serve many single antenna users. For a given number of users in a cell, the optimum number of active antennas maximizing EE has been derived where total BS downlink power is assumed to be fixed. Under the same assumption, the PAs have been dimensioned in a  way that maximizes network EE not only for a single time snapshot but over twenty four hours of operation while considering dynamic efficiency characteristics of the PAs. In order to incorporate this daily load profile, each BS has been modeled as an $M/G/m/m$ state dependent queue under the assumption that the network is  dimensioned  to serve a maximum number of users at a time corresponding to $100$\% cell traffic load. This load adaptive system along with the optimized PA dimensioning achieves $30$\% higher energy efficiency compared to a base line system where the BSs always run with a fixed number of active antennas which are most energy efficient while serving $100$\% traffic load.
\end{abstract}
\IEEEpeerreviewmaketitle

\begin{IEEEkeywords}

Massive MIMO;  Energy Efficiency; Power Amplifier Efficinecy; Dimensioning of Power Amplifier.

\end{IEEEkeywords}

\section{Introduction}
Network capacity demand is growing exponentially. The future generation of cellular network is required to be  energy efficient in order to reduce environmental impact as well as network operation cost while catering this capacity \cite{Da, Li}.  To design an energy efficient network, it is important to consider the fact that there is a significant variation of traffic demand throughout the day. The daily maximums are even 2-10 times higher than the daily minimums \cite{EarthModel}. However, today's cellular networks are designed and operated based on peak traffic demand without the capability to adapt according to the traffic variations from the Energy Efficiency (EE) perspective.  As a result, the energy consumption of a base station (BS) with no load is at least half of the maximum energy it consumes at the peak~\cite{EarthModel}. For $5$G cellular networks, it is very important to come up with systems and solutions that deliver very high capacity along with traffic adaptive operation capability so that energy consumption scales down proportionally with the network load. Massive MIMO has been touted to be    a leading candidate technology that can cater very high capacity. Under the massive MIMO system, "hundreds or even thousands of antennas simultaneously serve tens of user equipments (UEs) on the same time-frequency resource" \cite{Hoydis}.  This study aims to give an insight to the energy efficient design of massive MIMO system taking into account the dynamic efficiency of a power amplifier (PA)  and adaptive activation of antennas following a daily traffic profile. 

 Recently, PA aware MIMO capacity analysis has garnered significant attention.  
 In \cite{Vu}, the capacity for the single user MISO channel has been derived  considering per antenna power constraints when i) constant channel is known at both the transmitter and receiver and  ii) Rayleigh fading channel known only at the receiver. In \cite{Persson}, it has been shown that under the simultaneous per-antenna radiated power constraints and total consumed power constraints, the capacity achieving power allocation scheme feeds as many antennas as possible with maximum power. In~\cite{Hien}, it has been shown that transmit power can be reduced proportionally with the number of antennas if the BS has perfect channel state information (CSI) without any fundamental loss in performance. However, EE is shown to be a quasi-concave function with the number of antennas if the BS power model includes circuit power dissipated by analog devices and residually lossy factors~\cite{Ha}. In \cite{Emilc} and \cite{Emilj}, the authors show EE to be a quasi-concave function of each of the three parameters; namely, number of antennas, number of users and transmit power provided that the other two parameters are constant. They also conclude that contrary to the finding in \cite{Hien}, the energy optimal strategy requires  increasing the transmission power with the number of antennas if the circuit power consumption is considered.  However, in these MU-MIMO  studies, the impact of dynamic efficiency characteristics of the non-ideal  PAs and their energy optimal dimensioning based on daily load variation has not been considered.

 A PA reaches its maximum efficiency $\eta$ while operating at  its maximum output power, $P_{max,PA}$. However, due to the non-constant envelope signals, e.g., OFDM, CDMA,  the PAs rarely operate at their maximum output power. Usually the peak to average power ratio (PAPR) is around $8$  dB~\cite{wegener}. Therefore, $P_{max,PA}$ needs to be  around $8$ dB  higher than the maximum average output power of the  PA.  In case of  massive MIMO system, this dimension of the PA, i.e., $P_{max,PA}$  is a very important design parameter, especially, under the context of energy efficient operation of the network that experiences highly varying load throughout the day.

In this work, we consider the dynamic efficiency characteristics of the realistic PAs while designing a multi-cellular network where each BS  is equipped with a load adaptive massive MIMO system. With the assumption of fixed total downlink transmission power, we derive the optimum number of antennas, $M$ for any given number of users, $K$ and their combination related to the global maximizer of EE, denoted by $M_{gOpt}, K_{gOpt}$.  To find the optimal dimensioning of the PA that ensures $24$ hour energy efficient operation of the network, we need to capture the temporal load variation. In order to do that we derive the user distribution throughout the day by modeling each BS as a state dependent $M/G/m/m$ queue \cite{Cruz, Cheah} and utilize the daily load profile as suggested in~\cite{EarthModel, CLRL}. The $M/G/m/m$ queue dictates that for exponential arrival and general distribution of service time, maximum $m$ number of users can be served simultaneously (number of servers = $m$, waiting place = 0) and the state dependency arises from the fact that the user rate depends on the number of users in the system getting served simultaneously.  We assume the  maximum number of users that a  BS is allowed to serve is $m =K_{gOpt}$ and the network is dimensioned in way that these  $K_{gOpt}$ users constitute $100$\% load as per the daily load profile. Note that under this assumption the network becomes most energy efficient when serving maximum load.  The derived user distribution has been  used to formulate the optimum dimensioning problem for the PA that yields maximum EE over twenty four hour operation of the network.  We solve the formulated problem numerically to find the optimum maximum output power of the PA, $P_{max,PA}$. Note that we measure the EE in bit/Joule, i.e., ratio between the average achievable data rate and the total average power consumption~\cite{Emilc, Chen}. 

In our study, we consider both the traditional PA (TPA) and a more efficient PA, i.e., envelope tracking PA (ET-PA)~\cite{Hossain1,Hossain}. In case of the TPA, the maximum efficiency is achieved only when it operates  near the compression region, i.e., at maximum output power. On the other hand, ET-PA achieves better efficiency throughout the operating range by tracking the RF signal and regulating the supply voltage accordingly. Note that numerical analysis is used to derive the TPA results since there is no closed form expression to connect number of users and antennas.

From the numerical analysis, we observe that very high maximum output power of the PA  makes the BS energy efficient only when the number of users is small and very low output power results in higher efficiency only when the number of users is large. This leads to a clear choice of the maximum PA output power when daily load variation is considered. We also see that it is possible to increase network EE by 30\% with this load adaptive massive MIMO system along with optimally dimensioned PA compared to a base line network where the BSs always run with a fixed number of active antennas  realizing the highest EE when serving maximum cell load. 
 
The rest of the paper is organized as follows: 
In Section~\ref{sec:SystemModel}, we present the system model.
 In Section \ref{sec:ProblemFormulation},
we derive the load adaptive number of antennas and formulate the PA dimensioning problem. 
In Section \ref{sec:NumericalResults}, we illustrate the finding of the numerical analysis. We conclude the paper in Section~\ref{sec:Conclusion}.

\section{System model}
\label{sec:SystemModel}
In this work, we consider the downlink of a multi-cell massive MIMO system with $N$ number of cells where each cell has its own BS.  Each BS $c \in N$ transmits constant power, $P_c$  with $M_c$ number of antennas and serves  $K_c$ number of users equipped with a single antenna. We consider block flat-fading channels with $B_C$ (Hz) coherence bandwidth and $T_C$ (second)  coherence time which results  $U= B_C T_C$ static time-frequency blocks. Users are uniformly distributed in the regular hexagonal cell lattice. Typical 3GPP distance dependent pathloss model is used for large scale fading.  We consider perfect CSI and zero forcing precoding so that the intracell interference is canceled out by beam forming. Also, the power allocation is adapted to make sure that all the users achieve same data rate. Under these assumptions, the average data rate achieved by each user in cell $c$ is given by \cite{ Emilmo}

\begin{equation}
R_c= \beta \log(1+ \gamma_c (M_c-K_c))
\label{eq:Rc}
\end{equation}  								
where $\gamma_c=\frac{(P_c/K_c )}{(\Lambda_{cc} \sigma^2+ \sum_{d \neq c} \Lambda_{cd} P_d)}$ and  $\beta = (1-\frac{K}{T_c}) \frac{B}{\text{ln}2}$ , $T_c$ = coherence time,  $K$ = the maximum number of users  assumed to be the same for all cells, $B$ = Bandwidth,  $\Lambda_{cc} = \textrm{E}\{\frac{1}{\lambda_{serving}}\}$, where $\lambda_{serving}$ is the channel variance from the serving BS and  
$ \sum_{d \neq c} \Lambda_{cd} P_d$ = the average inter-cell interference power normalized by $\Lambda_1$.

\subsection{Power consumption model}                                  
Total power consumed in a BS can be given by 
\begin{equation} 
\label{eq:throughput}
P_c^{tot}= P_{PA} (P_c, M_c )+P_{BB} (P_c,M_c, K_c)+P_{Oth}
\end{equation}
where $P_{PA} (P_c,M_c )$ gives the power consumption in the PA when $M_c$ antennas transmit the total transmit power $P_c$ and $P_{BB} (P_c,M_c,K_c)$ is the base band processing  power where $K_c$ is the number of users served simultaneously. $P_{Oth}$ is the constant power consumption due to site cooling, control signaling etc. Here, power allocation to each antenna has been assumed to be equal.

\subsubsection{Power amplifier}
We consider TPA and ET-PA in this analysis. In case of TPA, the total input power needed for mean output  transmission power $p$ can be approximated as~\cite{Hossain} 
\begin{equation}\label{eq:T-PA}
P_{TPA}(p)\approx \frac{1}{\eta} \sqrt{p\cdot P_{max,PA}}
\end{equation}
where $\eta$ denotes the maximum PA efficiency and $P_{max,PA}$ is the maximum output power of the PA. Note  that, the maximum mean transmit power, $p_{max}$ must be around $8$ dB less than $P_{max,PA}$. 

In case of ET-PA, the total power consumption is approximately linear function of the actual transmit
power~\cite{Hossain},
\begin{equation} \label{eq:ET-PA}
P_{ET-PA}(p)\approx \frac{1}{(1+\alpha)\eta}\left(p+\alpha P_{max,PA}\right)
\end{equation}
where $\alpha\approx 0.0082$ is a PA dependent
parameter. 

\subsubsection{Baseband power consumption}
For baseband and fixed power consumption we use the model proposed in \cite{Emilc}. Total circuit power is given by 		
\begin{equation} 
\label{eq:Pcp}
P_{CP}=P_{TC}+P_{CE}+P_{(C/D)}+P_{LP} 	\nonumber
\end{equation}				
The power consumed in the transceiver is given by $P_{TC}=M P_{BS}+P_{SYN}$  where $P_{BS}$ is the power required to run the circuit components,  e.g., converters, mixers and filters attached to each antenna at the BS and $P_{SYN}$ is the power consumed by the local oscillator. $P_{CE}$ is the power required for channel estimation process. $P_{C/D}$ is the total power required for channel coding, $P_{COD}$  and channel decoding, $P_{DEC}$.   $P_{LP}$ is the power consumed for linear processing.
According to \cite{Emilc}, total baseband power can be expressed as 
\begin{eqnarray} 
\label{eq:Pbb}
P_{BB} (P_c,M_c,K_c)\!\!\! &=& \!\!\!\sum_0^3 C_{0,i}^{BB} K_c^i\!\!+\!\!M_c\sum_0^2 C_{1,i}^{BB} K_c^i\!\! +\!\! \mathcal{A} K_c R_c
\nonumber \\
&=& C_0^{BB} +M_c  C_1^{BB}   
\end{eqnarray}
where  $C_{0,0}=P_{SYN},  C_{0,1}=0,  C_{0,2}=0, C_{0,3}=\frac{B}{3UL_{BS}}, C_{1,0}= P_{BS}, C_{1,1}=\frac{B}{L_{BS}}(2+\frac{1}{U}), C_{1,2}=\frac{3B}{L_{BS}}, \mathcal{A}=P_{COD}+P_{DEC}$, $R_c$ is the rate achieved by a user on average as given in equation (\ref{eq:Rc}). 

In our study we define EE as following
 \begin{eqnarray}\label{EE}
EE &=&  \frac{\text{Average Sum Throughput}}{\text{Power Consumption}} 
\nonumber \\ 
 &=& \frac{K_c R_c}{P_{PA} (P_c, M_c )+P_{BB} (P_c,M_c,K_c )+P_{oth}}         
\end{eqnarray}
\subsection{Traffic model}  \label{TrafficModel}
We model the BS with massive MIMO system as a state-dependent $M/G/m/m$ queue  as in our setting we do not allow a BS to serve more than $K_{gOpt}$ number of users simultaneously and the users achieve different data rate at different state, i.e., when BS serves different number of users. Note that this $K_{gOpt}$, as derived in Section \ref{GlobalOpt}  is  the number of servers, $m$. We also assume the network to be  dimensioned in a way that while serving this maximum number of users, $K_{gOpt}$ the network load is highest, i.e., $100$\% corresponding to the daily load profile.  We allow at most $2$\% blocking at $100$\% load, i.e., the probability of serving  the maximum allowed number of users, $K_{gOpt}$ simultaneously is $0.02$.  In order to capture the daily traffic variation, we consider the  daily load profile proposed for data traffic in Europe  \cite{EarthModel} and the profile proposed for commercial and residential areas  \cite{ CLRL}.   The limiting probabilities for the random number of users $K$ in the $M/G/m/m$ state-dependent queueing model $\pi(n)\equiv Pr[K=n]$, are as follows \cite{ Cruz}
\begin{equation}
\pi(n)\!=\!\!\left[\frac{\left[\lambda\frac{\sigma}{R_1}\right]^n}{n! f(n)f(n-1)...f(2)f(1)} \right]\pi_0, n \!= \!1,2,...m,
\end{equation}
where $\pi_0$ is the probability that there is no user in the system and is given by
\begin{equation}
\pi_0^{-1}=1+ \sum_{i=1}^m\left(\frac{\left[\lambda\frac{\sigma}{R_1}\right]^i}{i! f(i)f(i-1)... f(2) f(1)}\right)  \nonumber
\end{equation}
where, $\frac{\sigma}{R_1}$ is the expected service time when there is a single user in the system, $f(n)=R_n/R_1$ where $R_n$
 is the  data rate per user while serving $n$ number of users, $\lambda$ is the arrival rate, and $\sigma$ is the total data traffic contribution by a single user. Note that we use (\ref{eq:Rc}) to find the data rates at different user states.

\begin{figure}[H] 
\centering
        \includegraphics[ height=1.8in]{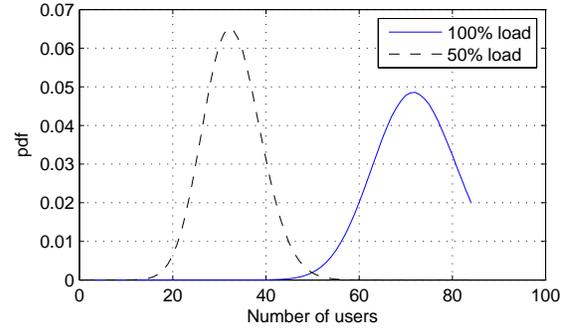}

\caption{User distribution while serving $50$\% and $100$\% cell load with the parameters given in Section \ref{sec:NumericalResults}.}
	\label{fig:Mgcc}
\end{figure}

 In order to find the steady state probability distribution throughout the day, first we set the values for $\lambda$ and $\sigma$. As we allow 2\% blocking rate  while serving $100$\% load, we find the maximum $\lambda$, i.e., $\lambda_{max}$ that results $\pi(K_{gOpt})=0.02$ for a fixed $\sigma$. Assuming that $\sigma$ remains constant, we derive the hourly average number of users following the daily load profile using $\lambda_{max}$. For example, from the load profile, if the load at any hour is  $x$\% , the corresponding average number of users is $\lambda_x= \frac{x}{100}\cdot \lambda_{max}$. Each of these hourly average number of users has been used as the input to the $M/G/m/m$ queue to find the steady state proabillity distribution  of the users during that hour. 
Figure \ref{fig:Mgcc} gives two example plots of the user distribution for 50\% and 100\% load with the parameters provided in Section \ref{sec:NumericalResults}. 
 
\section{Problem formulation}
\label{sec:ProblemFormulation}
 We solve two different problems in this study. Frist,  for a given number of users in a cell we find the optimum number of antennas that yields highest EE. We also find the global maximizer point corresponding to optimum number of users and number of antennas. In the second part we formulate the problem in order to find the optimum maximum output power of the PA, $P_{max, PA}$.

\subsection{Load adaptive number of antennas}
\label{GlobalOpt}

First, we reformulate the EE expression considering the dynamic efficiency characteristics of the PAs. Assuming that each antenna has its own PA and the total transmit power to be constant, the power per antenna is $P_c/M_c$  and the corresponding total input power for the PAs is 

\begin{equation}
P_{PA} (P_c,M_c )= \frac{M_c (P_c/M_c)} {\eta(P_c/M_c)} \nonumber
\end{equation}

where $\eta(P_c/M_c)$ gives the efficiency of the PA when it operates at the output power  $P_c/M_c$. Using the power model for ET-PA as in equation (\ref{eq:ET-PA})

\begin{eqnarray}
  P_{PA}(P_c,M_c) &=& M_c \frac{1}{(1+\alpha)\eta} (P_c/M_c +\alpha P_{max,PA})
	\nonumber \\
  &=&\frac{1}{(1+\alpha)\eta}P_c+ \frac{\alpha P_{max,PA}}{(1+\alpha)\eta} M_c \nonumber 
\end{eqnarray}

So, for both the baseband processing and PA, a part of the power consumption is constant and the other part of the consumption linearly increases with the number of antennas. Combining those parts accordingly total power can be written as  $P_{total} \approx C_0+C_1 M_c$  where $C_0=C_0^{PA}+C_0^{BB}+P_{Oth}$ and $C_1=C_1^{PA}+C_1^{BB}$.       
The EE in equation (\ref{EE}) can be written as 

\begin{equation}
\text{EE (ET-PA)} =\frac{K_c R_c}{C_0+C_1 M_c} \nonumber
\end{equation} 
The optimum number of antennas, $M_{c,Opt}$ for a given number of users  $K_c$ can be derived by using \cite{ Emilc} as follows
\begin{equation}
M_c=\frac{\exp \!\left[\mathcal{W}_0 \left[\frac{\gamma_c C_0-(1-K_c \gamma_c )C_1 )}{C_1 e}\right]\!+\!1\right]\!-(1\!-\!K_c \gamma_c)}{\gamma_c}
\end{equation}

where $\mathcal{W}_0$ is the main branch of the Lambert function.  In order to find the global optimum $EE (M_c, K_c)$, it is sufficient to calculate the optimum number of antennas and corresponding EE up to a reasonable number of users in the cell. Since  EE is a quasi-concave function of number  of users and number of antennas \cite{Emilc}, there is a global maximizer of  $EE (M_c, K_c)$ corresponding to the optimum number of antennas and users which is also a stationary point in the plane.  We denote these global maximizers, i.e., the optimum number of antennas and users by $M_{gOpt}$ and $K_{gOpt}$ respectively.

In case of TPA the EE equation in (\ref{EE}) takes the form 
\begin{equation}
\text{EE (TPA)} =\frac{K_c R_c}{C_0+C_1 M_c+ \sqrt{C_2 M_c}} \nonumber
\end{equation} 
For TPA, we do not have the closed form expression for the optimum number of antennas that maximizes the EE for a given number of users. However, as $M_c$ and $K_c$ are integer values, it is not computationally demanding to find the optimum number of antennas by exhaustive search for a given number of users. 

\subsection{Dimensioning of power amplifier}

As mentioned earlier, our focus is to design the massive MIMO system which results in an energy efficient network considering the load variation throughout the whole day. We allow a BS to serve  maximum $K_{gOpt}$ number of users simultaneously which has been derived in Section \ref{GlobalOpt}.  Using the $M/G/m/m$ queue, we find the distribution of the users over a day by using the hourly average number of users as described in Section \ref{TrafficModel}. Our objective is to design the system  in a way  that the EE weighted over this user distribution, as shown in (\ref{weightedEE}), gets maximized.
\begin {equation}
\text{EE (Weighted)} = \frac{1}{24} \sum_{h=1}^{24} {\sum_{n=1}^m {\pi(h, n) \cdot EE(n)}} 
\label{weightedEE}
\end{equation} 
where $\pi(h, n)$ is the steady state probability of state $n$,  i.e., probability that there are $n$ users in the system at any hour $h$.
As the number of antennas varies to cope with the varying load, power per antenna also varies in order to transmit fixed downlink power, $P_c$. Under this  circumstances,  there is a need to dimension the maximum output power, $P_{max,PA}$ for the PA that maximize $\text{EE (Weighted)}$. As the maximum average power a PA can transmit is 8 dB less than its  maximum output power, $P_{max,PA}$, for each $P_{max,PA}$ a certain number of PAs must remain always active in order to deliver $P_c$ W.  In order to achieve highest EE considering the daily load variation as well as dynamic efficiency characteristics of the PA, we need to solve the following problem:
\begin{subequations} 
\label{eq:problem}
\renewcommand\theequation{\theparentequation\roman{equation}}
\begin{align} 
   {\mathop {{\text{arg max}}\!\!:\!\!}_{P_{max,PA}}} &\;\;
   {\sum_{h=1}^{24}}
     \sum_{n=1}^m \!\!\pi(h, n) \frac{n R_c(n) }{C_0\!+\!(\!C_1^{BB} \!+\!C_1^* P_{max,PA}\!)M_c(n)} \label{eq:P11}\\
   {{\text{Subject to:}}} &\;\; {M_c(n)\geq n+1, P_{max,PA}\geq \frac{P_c}{M_c(n)}10^{0.8}} \label{eq:P12}  
\end{align}
\end{subequations}
 where $R_c(n)$ is the rate achieved by any user with the optimum number of antennas, $M_c(n)$ at the state, $K_c=n$ and  $ C_1^* = \frac{\alpha}{(1+\alpha)\eta}$. The first constraint comes from the prerequisite of zero forcing precoding, i.e., the number of antennas is larger than the number of users and the second constraint ensures that the minimum number of antennas required to transmit at least $P_c$ are active. 

We solve this problem using exhaustive search over the number of antennas. Given the total transmission power $P_c$, we can compute  the value of $P_{max,PA}$ for any number of active antennas, $M_c= 1,\ldots, M_{gOpt}$.  We end up with $M_{gopt}$ number of different values for $P_{max;PA}$ and select the one maximizing  (\ref{eq:P11}).

\section{Numerical analysis}
\label{sec:NumericalResults}

We consider the downlink of a cellular network with 19 regular  hexagonal cells where wrap around technique has been applied in order to get rid of the boundary effect. The total downlink power $P_c$ is constant at $20$ W and maximum cell radius is $500$ m if not specified otherwise. The users that resides in the inner $35$ m from the cell center are not considered in the analysis  and a uniform user distribution has been assumed outside that range. As the transmission power is fixed, we consider the interference to be fixed from neighboring cells and it is sufficient to derive the results for the central cell only. Note that we consider $15000$ test points in each cell in order to calculate average channel variance from the serving BS and the average inter-cell interference power. The parameters for the simulation as given in the following table are mainly taken from~\cite{Emilj}.

\begin{table}[!ht]
\caption{Simulation parameters}
\label{RP} \centering
\begin{tabular}{|l|l|} \hline
\multicolumn {2} {|c|}{\textbf {Reference parameters }} \\
 \hline
  \textit{Parameter} & \textit{Value} \\
 \hline
 Number of cells & $19$ \\
 \hline
Grid size inside each cell & $15000$ points \\
 \hline
Cell radius: $d_{max}$ & $500$ m \\
\hline
Minimum distance: $d_{min}$ & $35$ m \\
\hline
 BS average transmit power  & $20$ W\\
 \hline
Maximum PA efficiency  & $80$\%\\
 \hline
Path loss model & $\frac{10^{-3.53}}{||x||^{3.76}}$\\
 \hline
Local oscillator power: $P_{SYN}$  & $2$ W\\
\hline
BS circuit power: $P_{BS}$ & $1$ W \\
\hline
Other power: $P_{Oth}$  & $18$ W\\
\hline
Power for data coding: $P_{COD}$  & $0.1$ W/(Gbit/s) \\
\hline
Power for data decoding: $P_{DEC}$  & $0.8$ W/(Gbit/s) \\
\hline
Computational efficiency at BSs:$L_{BS}$ & 12.8 Gflops/W \\
\hline
Bandwidth  & $20$ MHz\\
 \hline
Total noise power: $B\sigma^2$  & -$96$ dBm\\
 \hline
Channel coherence time: $T_c$  & $10$ ms \\
 \hline
Channel coherence BW: $B_c$  & $180$ kHz \\
\hline
 Coherence block: $U$  & $1800$ \\
\hline
\end{tabular}
\end {table}

					\begin{figure}[!h] 
\centering
        \includegraphics[ height=2.0in]{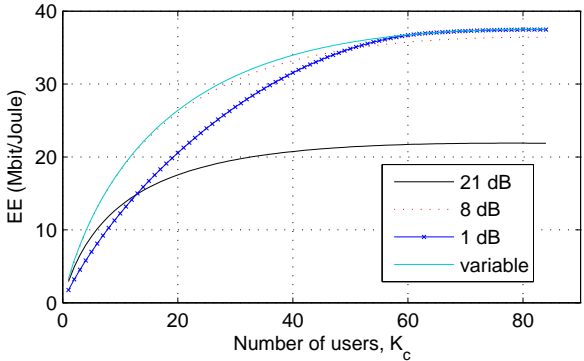}	
\caption{Impact of the dimensioning of $P_{max,PA}$ on EE at different cell loading (ET-PA)}
\label{fig:ExamplePmax}
\end{figure}

					\begin{figure}[!h] 
\centering
        \includegraphics[ height=2.0in]{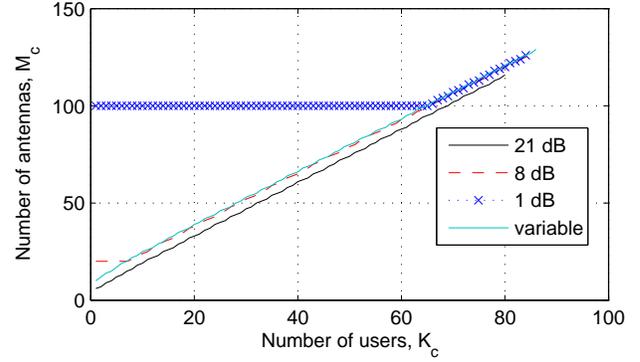}	
\caption{Required number of antennas, $M_c$ for a given number of users under different $P_{max,PA}$ maximizing EE.}
\label{fig:OnlyAntennas}
\end{figure}	
	
 \subsection{Optimum maximum output power of PA, $P_{max,PA}$}
 \label{OptPmaxPA}

We illustrate the impact of different maximum output power of the PA, i.e., $P_{max,PA}$ on EE in  Figure~\ref{fig:ExamplePmax} and  the corresponding number of antennas required to achieve the maximum EE for a given number of users in Figure~\ref{fig:OnlyAntennas}.  As the total average transmit power $P_c$ is fixed, the transmit power per antenna, $P_c/M_c$ varies with the number of antennas. For the 'variable' curve, we do not restrict the PA with a fixed  $P_{max,PA}$ rather it has been allowed to be flexible, i.e., just 8 dB higher than the average transmit power of the PA and hence gives the upper bound of EE. In other cases,  $P_{max,PA}$ is fixed disregarding the actual transmit power transmitted by each antenna. However, it was ensured that minimum number of antennas was always active in order to guarantee fixed transmission power.  For example, '$8$ dB' plot represents the case when at least $20$ antennas are required to deliver total transmit power, $P_c= 20$ W, see Figure~\ref{fig:OnlyAntennas}. 

\begin{figure}[t]
\centering
        \includegraphics[ height=2.0in]{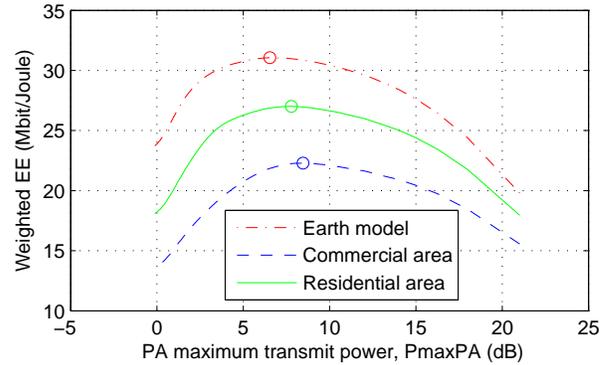}	
\caption{Impact of load variation pattern on the design of PA while using ET-PA.}
\label{fig:OptimumPmax}
\end{figure}
\begin{figure}[t]
\centering
        \includegraphics[ height=2.0in]{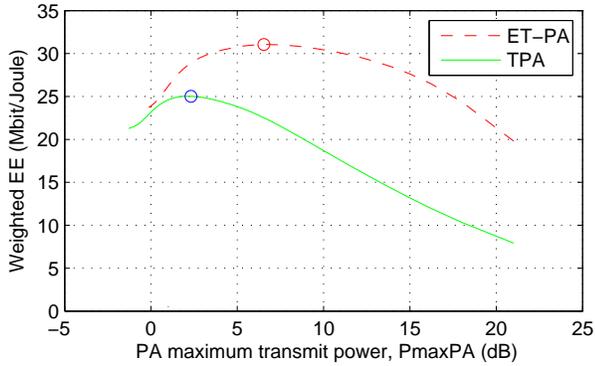}
\caption{EE performance comparison between TPA and ET-PA.}
	\label{fig:RCEarth}
\end{figure}
It can be seen from these figures that if $P_{max,PA}$ is very high, e.g., $21$ dB, it is comparatively efficient when the number of users are pretty small as it provides the option to run the BS even with a single antenna. When the number of users is large the minimum number of active antennas is also large as it needs to be larger than the number of users. This leads to low output power per antenna and renders the system inefficient because the PA operates away from the compression region. On the other hand, when the $P_{max,PA}$ is low, e.g., '$1$ dB', it becomes very efficient at high load as addition of each antenna contributes little in terms of power but increases array gain. However, it is not much efficient for smaller number of users as the number of active antennas cannot be small due to total transmit power constraint. Note that the PA with $1$ dB $P_{max,PA}$ delivers $0.20$ W transmit power, hence at least $100$ of them are always active to deliver $20$ W. Another observation from Figure~\ref{fig:OnlyAntennas} is that, with higher $P_{max,PA}$ the number of antennas required to achieve maximum EE for a given number of users is smaller compared to the case of lower $P_{max,PA}$. Also, the maximum number of users that achieve global maximum EE, i.e., $K_{gOpt}$ is smaller in case of higher $P_{max,PA}$. Based on Figure 3, one can see that the ratio between the optimal number of antennas in terms EE is around twice compared to  the number of users, i.e., $M/K \approx 2$. 

As our objective is to design the network that is energy efficient not only in the busy hour but also throughout the whole day, in Figure~\ref{fig:OptimumPmax}, we show the impact of different fixed $P_{max,PA}$ on EE while weighted over daily load variation. For the daily load variation we use both the model proposed for data traffic in Europe and the model proposed for commercial and residential areas \cite{EarthModel, CLRL}. Note that there is an optimum value for $P_{max,PA}$ that makes the network most energy efficient. Also, for a wide range of $P_{max,PA}$, the EE is close to the maximum achievable by the optimum $P_{max,PA}$. However, as shown in Figure~\ref{fig:RCEarth}, in case of TPA, the efficiency reduces drastically if the $P_{max,PA}$ drifts away from the optimum value. Another observation is that the difference between daily load profile of commercial and residential areas does not have significant impact on the dimensioning of the PA.

\begin{figure}[t]
\centering
        \includegraphics[ height=2.0in]{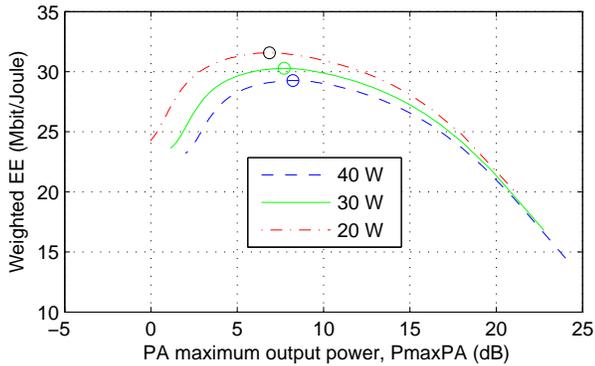}
\caption{Impact of different power level (fixed) on the design of PA while using ET-PA.}
	\label{fig:Pcvariable}
\end{figure}
				
\begin{figure}[t]
\centering
        \includegraphics[ height=2.0in]{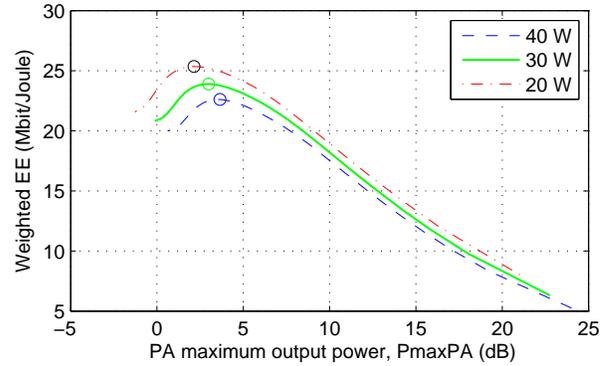}
\caption{Impact of different power level (fixed) on the design of PA while using TPA}
	\label{fig:TPAPcvariable}
\end{figure}
\subsection{Impact of cell size and power level on design of PA}
The impact of cell size on the dimensioning of PA was found negligible. However, the impact of the level of total transmission power is more prominent as shown in Figure~\ref{fig:Pcvariable} and Figure~\ref{fig:TPAPcvariable}. We see that   the difference in EE with different power level is comparatively less significant in case of ET-PA.  It is also evident from the figures that  in case of TPA, the optimum $P_{max,PA}$ is much lower compared to its counter-part, ET-PA.  This is  due to the fact that, the efficiency of a TPA falls drastically if the output power drifts from the maximum. As a result, systems with TPAs requires  PAs  with smaller maximum power so that more PA's can be pushed towards compression region where the efficiency is very high. Consequently, the number of users and the number of antennas required to achieve the global optimum EE are also higher in case of TPA.

\begin{figure}[!t]
\centering
        \includegraphics[ height=2.0in]{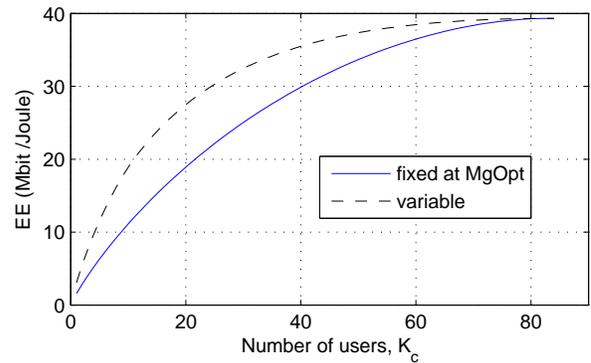}
\caption{Comparison of EE between proposed load adaptive system and fixed base line system for all user states.}
	\label{fig:AA}
\end{figure}

\balance
\subsection{Benefit of load adaptive antenna system with PA dimensioning}

In Figure \ref{fig:AA}, we illustrate the performance of the load adaptive antenna system compared with a base line system where the number of antennas are fixed and optimized according to the maximum cell load, i.e., the number of antennas that maximize the EE during 100\% cell loading are always active disregarding the number of users the BSs serve.  We consider the default values for the parameters, e.g., total downlink power, $P_c = 20$ W, and inter-site distance, i.e., cell radius $d_{max} = 500$ m. Also, we consider the PA that we have dimensioned in Section \ref{OptPmaxPA} where the corresponding $P_{max,PA}$  was found to be 4.485 W. The plot 'variable' indicates the EE achieved by this load adaptive antenna system with the optimally dimensioned PA  and the plot 'fixed at $M_{gOpt}$' represents the EE achievable if the system always run with the fixed number of antennas, $M_{gOpt}$ which is the optimum number maximizing the network  EE at the maximum cell load. It is evident from the figure that there is significant increase in EE with the adaptive antenna for almost all the user states. We use the weighted EE formula as given in  (\ref{weightedEE}) utilizing the daily load profile as proposed in \cite{EarthModel} and found that the overall increase in EE considering twenty four hour load variation is 30\%.

\section{Conclusion}
\label{sec:Conclusion}
In this study, we design a multicellular network with load adaptive massive MIMO system and investigate the impact of daily load profile on the design of PA and number of active antennas. Modeling a BS with massive MIMO as a $M/G/m/m$ queue, we generate the user distribution following the daily load profile and show that  daily load variation has significant impact on the design of the PAs. It  renders PAs with very high or very low output power inefficient and gives a clear choice for the optimum value. From the comparison between the performance of TPA and ET-PA, it has been found that the ETPA  with better efficiency  over wider operating range not only yields higher EE but also provides more flexibility to design the PA. Numerical analysis also suggests that this load adaptive MIMO system along with optimally dimensioned PA increases network EE about 30\% even when compared to a base line network where the BSs run with a fixed number of antennas that yield highest EE at $100$\% cell loading.

\section*{Acknowledgment}

This work has been  prepared in the frameworks of EWINE-S project  supported by the Finnish Funding Agency for Technology and Innovation (TEKES) and 5grEEn, an EIT-ICT Labs project  funded by the EIT, European Institute of Innovation and 
Technology.  The authors would  also like to thank Mr. Emil Bj\"{o}rnson for his remarks.

\end{document}